\documentstyle{laa}

\begin{document}
\thesaurus{07  
	   (07.09.1; 
	    07.13.1;  
	   )}

\title{Solar Wind and Motion of Meteoroids}

\author{ Jozef Kla\v{c}ka }
\institute{Institute of Astronomy,
Faculty for Mathematics and Physics,
Comenius University, \\
Mlynsk\'{a} dolina,
842 15 Bratislava,
Slovak Republic}
\date{}
\maketitle

\begin{abstract}
The effect of nonradial component of solar wind is discussed from
the qualitative point of view. It is shown that the direction
of nonradial component is opposite in comparison with the direction
used in papers dealing with orbital evolution of meteoroids.
\end{abstract}

\section{Introduction}
Kla\v{c}ka (1994) has discussed the effect of solar wind particles on the
motion of small interplanetary dust particles. He has pointed out that
nonradial component of the solar wind has an opposite direction than the
direction used in papers of the field of interplanetary matter. He has also
tried to obtain some quantitative results following from consideration
of the correct nonradial direction.

One of the newest papers dealing with the orbital evolution of meteoroids
(Cremonese {\it et al.} 1997)
also takes nonradial direction of solar wind into account. However, this
direction is still taken in the incorrect way (see also references in
Cremonese {\it et al.} 1997).

\section{Nonradial Direction of the Solar Wind -- Qualitative Discussion}
Using literature on solar physics, Kla\v{c}ka (1994) has pointed out
that nonradial component of solar wind brings down the effect
of inspiralling interplanetary dust particles toward the Sun generated
by the effect of solar electromagnetic radiation (spherical particles:
Robertson 1937,
see also Kla\v{c}ka 1992 as for the most complete form;
nonspherical particles: Kla\v{c}ka 1994, Kla\v{c}ka and Kocifaj 1994).
However, according to
(Cremonese {\it et al.} 1997, see also references in this paper)
the nonradial component of the solar wind raises the effect
of inspiralling interplanetary dust particles toward the Sun generated
by the effect of solar electromagnetic radiation.

We can present another argument in favour of the nonradial direction
of solar wind used in Kla\v{c}ka (1994). This argument is presented in
Figure 3.20 (p. 84) in Hundhausen (1972). The important property of the
nonradial component deals in fact that it is a function of distance from
the Sun (e. g., Figure 3.21 (p. 87) in Hundhausen (1972)). This fact was
not considered in Kla\v{c}ka (1994). However, the dependence on distance
is not known, and, thus, no precise calculations can be reliably done.
The only thing one can done is to make some other estimates
(in comparison to Kla\v{c}ka 1994), using as a
motivation the result presented in Figure 3.21 (Hundhausen 1972, p. 87).
Supposing that the nonradial component depends on the heliocentric
distance as $1/r$, the inspiralling toward the Sun due to the
solar electromagnetic radiation will be stopped at several solar radii.
This artificial example shows the importance of knowing exact dependence
of the nonradial component on the heliocentric distance.



\section{Conclusion}
We have shown that solar wind's real direction of nonradial component
is opposite in comparison to that considered in
(Cremonese {\it et al.} 1997, see also references in this paper).
Correct consideration of this nonradial component, together with its
dependence on heliocentric distance, will help us in our better
understanding of the evolution of small particles in solar system
(perhaps one of the most interesting concerns the zodiacal cloud).

%
%
%
%
%


\end{document}